\documentclass[pra,twocolumn,amsfonts,amssymb,amsmath,]{revtex4}

\usepackage{epsfig}

\newcommand{\be}{\begin{eqnarray}}
\newcommand{\ee}{\end{eqnarray}}
\newcommand{\bm}{\boldsymbol}

\newcommand{\n}{\nonumber}

\newcommand{\fD}{~_0D^{1-\alpha}_t}

\begin{document}

\title{Fractional Fokker-Planck equations for subdiffusion\\
and exceptional orthogonal polynomials}

\author{C.-L. Ho}
\affiliation{Department of Physics, Tamkang University,
 Tamsui 251, Taiwan, R.O.C.}


\begin{abstract}

It  is pointed out that, for the fractional Fokker-Planck equation for subdiffusion proposed by Metzler, Barkai, and Klafter [Phys. Rev. Lett. 82 (1999) 3563], there are four types of infinitely many exact solutions associated with the newly discovered exceptional orthogonal polynomials.  
They represent fractionally deformed versions of the Rayleigh process and the Jacobi process. 

\end{abstract}




 \maketitle




\section{\bf Introduction}

In recent years anomalous diffusions have attracted much interest owing to their ubiquitous appearances in many physical situations.
 For examples, charge carrier transport in amorphous semiconductors, nuclear magnetic resonance diffusometry in percolative, and porous systems, Rouse or reptation dynamics in polymeric systems, transport on fractal geometries, and many others \cite{MK}.
Unlike the  well-known Brownian motion, anomalous diffusions are  characterized by a mean-squared displacement relation   $\langle x^2 \rangle\sim t^\gamma(\gamma\neq 1)$ which is not linear in time --  it is  superdiffusive for $\gamma>1$, and subdiffusive for $\gamma<1$.

   The Brownian motion is Markovian in nature.  This means each new step in the motion depends only on the present state, and is independent of the the previous states.  Anomalous diffusion arises, on the contrary, from some memory effect of previous states, or as a result of some fractal structure of the background space, or due to some non-linear interaction inherent in the system, etc.  Among various ways to model anomalous diffusions, one interesting and successful way is by fractional differential equations that involve derivatives of fractional order [2-19].  We note here that a type of superdiffusive motion,  so-called the L\'evy flight, has been modeled by a diffusion equation with the Riesz-Weyl type fractional space-derivative \cite{D3,D4}, and the subdiffusive diffusion has been described  by  fractional Fokker-Planck equations (FFPE) \cite{FP1,FP2,FP3,FP4,FP5}.
   
It is natural that one would seek exact solutions of these fractional stochastic equations.  However, as is often the case, exact solutions are hard to come by.   The purpose of this note is to point out that  for the FFPE proposed in \cite{FP1,FP2,FP3}, there are four types of infinitely many exact solutions associated with the newly discovered exceptional orthogonal polynomials.  This is a direct extension of the results in \cite{CH}.



\vskip 0.5cm

\section{Fractional FPE
 and Schr\"odinger equations}

The FFPE proposed in \cite{FP1,FP2,FP3} is
\be
\frac{\partial}{\partial t} P(x,t)&=&\fD \left[-\frac{\partial}{\partial x} D_1(x)+
\frac{\partial^2}{\partial x^2}D_2\right]P(x,t),\n\\
&& ~~0<\alpha<1.
 \label{FFP}
\ee
Here $P(x,t)$ is the probability density function (PDF) , and 
$D_1(x)$ and $D_2={\rm\ constant}$ are, respectively,  the fractional drift and diffusion
coefficients. 
$\fD$ is the Riemann-Liouville fractional derivative defined by \cite{B1,B2}
\be
\fD f(x,t) =\frac{1}{\Gamma(\alpha)}\frac{\partial}{\partial t} \int^t_0 (t-t^\prime)^{\alpha-1} f(x,t^\prime)\, dt^\prime\n
\label{RL}
\ee
for $0<\alpha<1$.
The general situation  where the drift and diffusion coefficients are space-time dependent is considered in \cite{FP4,FP5}.

Without loss of generality, we set  $D_2=1$.   Now let
\be
P(x,t)=T(t)\,\Phi(x),
\ee
then we have
\be
\frac{d}{dt}\,T(t)=- \mathcal{E} \fD T(t),\label{FFP1}\\
\left[-\frac{\partial}{\partial x} D_1(x)+
\frac{\partial^2}{\partial x^2}\right]\Phi(x)=-\mathcal{E}\Phi(x).\label{FFP2}
\ee
Solution of the temporal part  (\ref{FFP1}) is given by \cite{MK,FP1}
\be
T(t)=E_\alpha (-\mathcal{E}\,t^\alpha),
\ee
where $E_\alpha (z)$ is the Mittag-Leffler function \cite{B1,B2}
\begin{equation}
E_\alpha(z)=\sum_{k=0}^\infty \frac{z^k}{\Gamma(\alpha k +1)}, ~~\alpha\in {\mathbb C}, ~~{\rm Re}\  \alpha >0.
\end{equation}

The drift coefficient can be defined by a
drift potential $U(x)$ as $D_1(x)=-U^\prime (x)$, where the prime denotes derivative with respective to $x$. 
Substituting
 \be
\Phi(x)\equiv  e^{-U(x)/2} \phi(x)
 \ee
 into Eq.\,(\ref{FFP2}), 
one finds that $\phi$ satisfies the  time-independent Schr\"odinger 
equation $H\phi=\mathcal{E}\phi$ with Hamiltonian $H$ \cite{CH,Ris}
\begin{eqnarray}
H=-\frac{\partial^2}{\partial x^2}+\frac14\, U^\prime (x)^2 -
\frac12\, U^{\prime\prime}(x).\nonumber
\end{eqnarray}
 and eigenvalue $\mathcal{E}$. 
$\phi_0=\exp(-U(x)/2)$ is the zero mode of $H$: $H\phi_0=0$.

Thus the FFPE  can be exactly solved if the corresponding Schr\"odinger equation is exactly solvable.
One needs only to link the $U(x)$ in the Schr\"odinger
system with the drift
coefficient $D_1(x)=-U^\prime (x)$ in the FFPE.  
Particularly, those related to the exceptional orthogonal polynomials discussed in \cite{CH} can be carried over directly, and hence an infinitely number of exactly solvable FFPE are found.

Let $\mathcal{E}_n$ and $\phi_n$ ($n=0,1,2,\ldots$) be the eigenvalues and 
eigenfunctions  of $H$, then the solution of Eq.\,(\ref{FFP}) is \cite{CH,MK,FP1}
\begin{equation}
P(x,t)=\phi_0(x)\sum_n c_n \phi_n(x) E_\alpha (-\mathcal{E}_n\,t^\alpha).
\label{pdf}
\end{equation}
with constant coefficients $c_n$ ($n=0,1,\ldots$)
\begin{eqnarray}
c_n=\int_{-\infty}^\infty \phi_n(x)\left(\phi_0^{-1}(x)
P(x,0)\right)dx.
\label{c_n}
\end{eqnarray}
Note that  $c_0= \int_{-\infty}^\infty 
P(x,0)dx=1$.
The stationary distribution, corresponding to $\mathcal{E}_0=0$,  is $P_0(x)=\phi_0^2(x)=\exp(-U(x))$ (with $\int
P_0(x)\,dx=1$), which is obviously  independent of $\alpha$.

In this paper we shall be interested in the  initial profile given by the $\delta$-function
\begin{equation}
P(x,t)=\delta(x-x_0),
\end{equation}
which corresponds to the situation where the particle performing the stochastic motion is initially located at the point $x_0$.
From Eqs.~(\ref{c_n}) and (\ref{pdf}) we have
\begin{equation}
c_n=\phi_0^{-1}(x_0)\phi_n(x_0),
\end{equation}
and
\begin{equation}
P(x,t)=\phi_0(x)\phi_0^{-1}(x_0)\sum_n \phi_n (x_0) \phi_n(x) E_\alpha (-\mathcal{E}_n\,t^\alpha).
\label{pdf-d}
\end{equation}

The fractional Ornstein-Uhlenbeck process with $U(x)=x^2/2$ have been discussed in \cite{MK,FP1}.  Here we would like to extend the solvable cases to those related to the singled-indexed exceptional orthogonal polynomials, which turn out to be fractional generalisation of the Rayleigh process \cite{Ray} and the Jacobi process \cite{KT}.


\section{FFPEs with Exceptional orthogonal polynomials}

 Discoveries of the so-called exceptional orthogonal polynomials, and the
 quantal systems related to them, have been among the most interesting developments in mathematical physics
in recent  years [24-38].  Unlike the classical orthogonal polynomials, these new polynomials have the remarkable
properties that they  start with degree $\ell=1,2\ldots$
polynomials instead of a constant, and yet they still 
 form complete sets with respect to some positive-definite measure.  The first new polynomials are associated with a single index $\ell$, but soon new polynomials with multiple indices were discovered.   For a review of these polynomials, please see e.g., 
Ref.\,\cite{Que5,Sasaki,UM}.

The new polynomials that fit in the construction described in the last section are the single-indexed ones. 
There are four basic families of the single-indexed exceptional orthogonal polynomials: two associated with the Laguerre types 
(termed the L1 and L2 types) ,
 and the other two with the Jacobi types (termed the J1 and J2 types).
Quantal systems associated with the Laguerre-type exceptional orthogonal polynomials are related to the radial oscillator, 
while the Jacobi-types are related to the  Darboux-P\"oschl-Teller potential.
The corresponding FPEs describe the
generalized Rayleigh process, and the generalized 
Jacobi process, respectively.

The eigenfunctions of the corresponding Schr\"odinger equations 
are given by 
\begin{equation} 
\phi_{\ell,n}(x;\bm{\lambda})=N_{\ell,n}(\bm{\lambda})\phi_{\ell}(x;\bm{\lambda})P_{\ell,n}(\eta(x);\bm{\lambda}),
\end{equation}
where $\eta(x)$ is a function of $x$, $\bm{\lambda}$ is a set of parameters of the systems, $N_{\ell,n}(\bm{\lambda})$ is a normalization constant, 
$\phi_{\ell}(x;\bm{\lambda})$ is the asymptotic factor, and $P_{\ell,n}(x;\bm{\lambda})$ is the polynomial part of the wave function,
which is given by the exceptional orthogonal polynomials.
When $\ell=0$, $P_{0,n}(x;\bm{\lambda})$ are simply the ordinary classical orthogonal polynomials of the corresponding quantal systems, 
with $P_{0,0}(x;\bm{\lambda})={\rm ~constant}$.  When $\ell>0$, $P_{\ell,n}(x;\bm{\lambda})$ is a polynomial of degree $\ell+n$. 
The PDF  (\ref{pdf-d}) for these generalized systems are
\begin{eqnarray}
&& P(x,t)=\phi_\ell^2(x;\bm{\lambda}) P_{\ell,0}(\eta;\bm{\lambda}) P^{-1}_{\ell,0}(\eta_0;\bm{\lambda})\\
&&\times  \sum_{n=0}^\infty N_{\ell,n}(\bm{\lambda})^2 P_{\ell,n}(\eta;\bm{\lambda})P_{\ell,n}(\eta_0;\bm{\lambda})E_\alpha (-\mathcal{E}_n\,t^\alpha).\n
\label{pdf-d-ell}
\end{eqnarray}

Infinite number of exactly solvable FFPE's can be constructed in terms of these four sets if new polynomial, using the basic data of these four systems summarized in \cite{CH}. 

Here we shall only illustrate the construction with the L1 Laguerre polynomials.

The function $U(x)$  that generates the  original radial
oscillator potential  is
\be
U_0(x;g)=x^2 - 2g\log x,~~0<x<\infty.
 \label{w0}
 \ee
The exactly solvable Schr\"odinger equations associated with the deformed radial oscillator potential 
are defined by  ($\ell=1,2,\ldots$)
\begin{equation}
U_\ell(x;g)= x^2 - 2(g+\ell)\log x
-2 \log\frac{\xi_\ell(\eta;g+1)}{\xi_\ell(\eta;g)}.\label{wl-osc}
\end{equation}
Here  $\bm{\lambda}=g>0$ and 
$\eta(x)\equiv  x^2$ and $\xi_\ell(\eta;g)$ is a deforming function, which for $L1$ type is given by
 \be
  \xi_{\ell}(\eta;g)=
  L_{\ell}^{(g+\ell-\frac32)}(-\eta).
  \label{xiL}
\ee 
The eigen-energies are
$\mathcal{E}_{\ell,n}(g)=\mathcal{E}_n(g+\ell)=4n$, which
are independent of $g$ and $\ell$. Hence the deformed radial
oscillator is iso-spectral to the ordinary radial oscillator. The
eigenfunctions are given by
 \be
\phi_{\ell,n}(x;g) &=&N_{\ell,n}(g) \phi_\ell(x;g) P_{\ell,n}(\eta;g),~~~\\
\phi_\ell(x;g)&\equiv & \frac{e^{-\frac{1}{2}
x^2}x^{g+\ell}}{\xi_{\ell}(x^2;g)}, \label{phi-l}
\ee
where the corresponding exceptional Laguerre polynomials
$P_{\ell,n}(\eta;g)$ ($\ell=1,2,\ldots$, $n=0,1,2,\ldots$) can be
expressed as a bilinear form of the classical associated Laguerre polynomials
$L^\alpha_n(\eta)$ 
and the deforming polynomial $\xi_\ell(\eta;g)$, as given in \cite{HOS}:
\be
P_{\ell,n}(\eta;g)&=&
  \xi_{\ell}(\eta;g+1)P_n(\eta;g+\ell-1)\n\\
 && -\xi_{\ell}(\eta;g)\partial_{\eta}
  P_n(\eta;g+\ell-1)  \label{XL},
\ee
where
$P_n(\eta;g)\equiv
  L_n^{(g-\frac{1}{2})}(\eta)$.
 The polynomials
$P_{\ell,n}(\eta;g)$ are degree $\ell+n$ polynomials in $\eta$ and
start at degree $\ell$: $P_{\ell,0}(\eta;g)=\xi_{\ell}(\eta;g+1)$.
They are orthogonal with respect to certain weight functions,
which are deformations of the weight function for the Laguerre
polynomials (for details, see \cite{HOS}).
From the orthogonality relation of $P_{\ell,n}(\eta;g)$, we get the normalization constants \cite{HOS}
\begin{equation}
  N_{\ell,n}(g)=
  \left[\frac{2n! (n+g+\ell-\frac12)}{(n+g+2\ell-\frac12)\Gamma (n+g+\ell+\frac12)}\right]^{\frac12}.
\end{equation}

The drift coefficient and PDF of the FFPE  generated by the drift potential $U_\ell(x)$ in Eq.\,(\ref{wl-osc}) is
\be
&&D^{(1)}(x) =-2x+ 2\frac{g+\ell}{x}\n\\
&&~~~~~ +4x\left(\frac{\partial_\eta\xi_\ell(\eta;g+1)}{\xi_\ell(\eta;g+1)}-\frac{\partial_\eta\xi_\ell(\eta;g)}{\xi_\ell(\eta;g)}\right)
\label{D1-L}
\ee
and
\begin{eqnarray}
&& P(x,t)=\phi_\ell^2(x;g) P_{\ell,0}(x^2;g) P^{-1}_{\ell,0}(x_0^2;g)\label{pdf-L}\\
&& \times  \sum_{n=0}^\infty N_{\ell,n}^2 P_{\ell,n}(x^2;g)P_{\ell,n}(x_0^2;g) E_\alpha (-4 n\,t^\alpha).\n
\end{eqnarray}

In the limit $\alpha\to 1, \ell\to 0$ for $g=1/2$,  we have $\xi_\ell(x;g+1)\to 0$ and $\xi_\ell(x;g)\to 0$, and eqs.~(\ref{D1-L}) and (\ref{pdf-L}) reduce to the drift coefficient and PDF for the ordinary Rayleigh process, 
\be
D^{(1)}&=&-2x+\frac{1}{x},\\
P(x,t)&=&2\, x\, e^{-x^2}\sum_{n=0}^\infty L_n(x^2) L_n(x_0^2)  \exp(-4n t),\n
\ee
where $L_n(x)$ are the ordinary  Laguerre polynomials.

\bigskip

\section{Numerical results}

In \cite{CH} we have studied how the Rayleigh process was deformed by $\ell$.  Here the effect of different $\alpha$ on the process  is considered numerically for the original Rayleigh process ($g=1/2, \ell=0$ ), and the $\ell$-deformed one ($g=1/2, \ell=5$). The choice of $\ell=5$ is due to an observation noted in \cite{CH}, explained below.  

The drift coefficient is related to the negative slope of $U_\ell(x)$ by $D_2(x)=-U_\ell^\prime(x)$.
So one  can gain some qualitative understanding of how the peak of the probability density moves just from the sign of the
slope of $U_\ell(x)$.
The peak of $p(x,t)$  tends to move to the right (left) when it is at a position $x$ such that $U_\ell^\prime(x)$ is
negative (positive), until the stationary distribution is reached.

 Fig.~1 depicts $U_\ell(x)$ for $g= 0.5$ and  $\ell=0$ (the original Rayleigh process), $1$ and $5$.  
 Note that  there can be
sign change of the slope of the drift potential  in certain region
near the left wall. In such region,  the peak of the probability density function will move in different directions for different  $\ell$. Particularly, 
at $x_0=1.2$, the sign of the slope of $U_5(x)$ is different from those of  $U_0(x)$ and $U_1(x)$. Thus one expects that
for the initial profile $P(x,0)=\delta(x-x_0)$ with the peak initially located at $x_0=1.2$, the peak will move to the right for $\ell=5$ system, while for the other 
two values of $\ell$, the peak will move to the left.   

Fig.\,2 and 3 show how $\alpha$ affects the Rayleigh process ($g=1/2, \ell=0$ ) and the $\ell$-deformed process ($g=1/2, \ell=5$), respectively, .
As expected,  the peak moves to the left for the Rayleigh process, and to the right for the $\ell$-deformed one.  Further, it is clear from the graphs that for $0<\alpha<1$ the FFPE describes subdiffusive motion -- the peaks for $\alpha\neq 1$ move slower to the left that the Rayleigh motion in Fig.\,2,  while they move faster to the right in Fig.\,3.  And the smaller the value of $\alpha$, the slower the peak moves.  From the last sub-figure it is obvious that  the PDF at large times is the stationary distribution which is independent of $\alpha$.




\section{Summary}

In this note we have discussed the construction of four types of infinitely many exact solutions, associated with the newly discovered exceptional orthogonal polynomials, for the fractional Fokker-Planck equations for subdiffusion proposed in \cite{FP1,FP2,FP3}.
They represent the fractionally deformed versions of the Rayleigh process and the Jacobi process.








\onecolumngrid


\begin{figure}[ht] \centering
\includegraphics*[width=15cm,height=6cm]{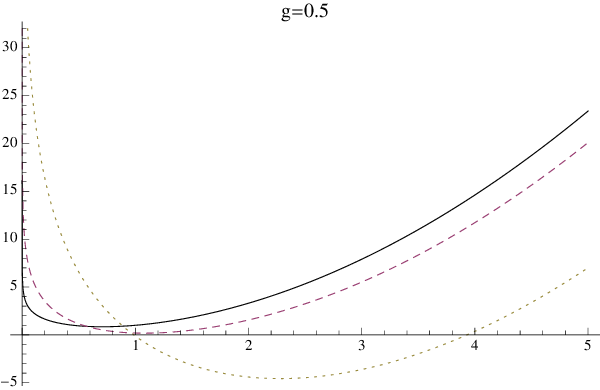}
\caption{Plot of $U_\ell (x)$ versus $x$ for the L1 type FFPE with  $g=1/2$ and $\ell=0$ (solid curve), $1$ (dashed curve) and $5$ (dotted curve) }
\label{Fig1}
\end{figure}


\onecolumngrid

\begin{figure}[ht] \centering
\includegraphics*[width=8cm,height=6cm]{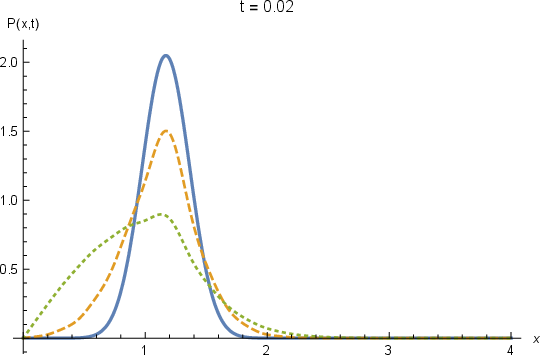}\hspace{1cm}
\includegraphics*[width=8cm,height=6cm]{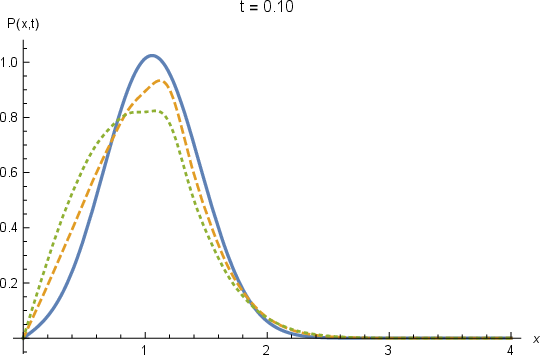}\\ 
~\\
~\\
~\\
\includegraphics*[width=8cm,height=6cm]{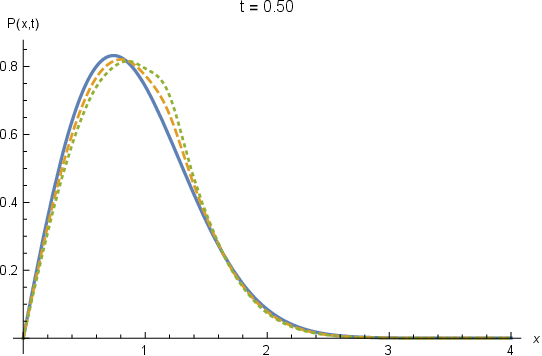}\hspace{1cm}
\includegraphics*[width=8cm,height=6cm]{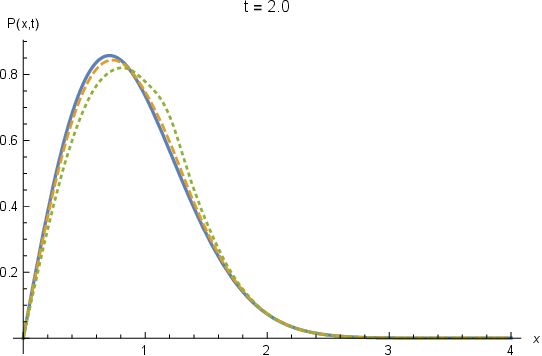}
~\\
~\\
~\\
\includegraphics*[width=8cm,height=6cm]{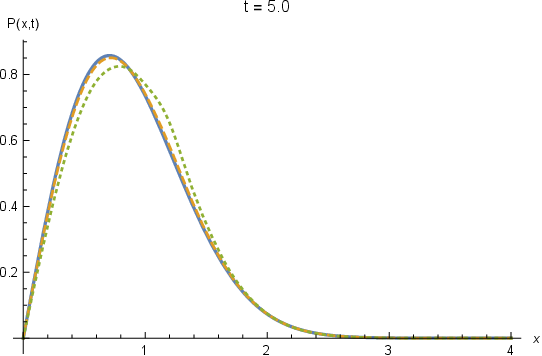}\hspace{1cm}
\includegraphics*[width=8cm,height=6cm]{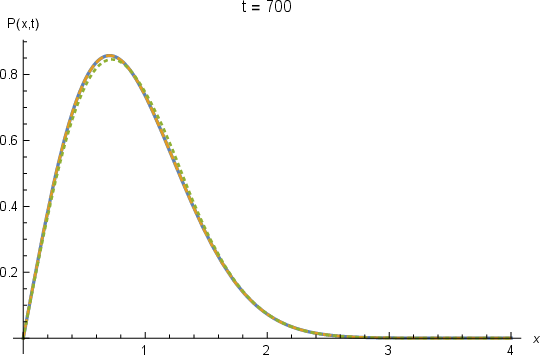}
\caption{Plots of $P(x,t)$ at different times for the Rayleigh process  
with $g=0.5, \ell=0, x_0=1.2$, 
$\alpha=1$ (solid curve), $0.75$ (dashed curve) and $0.25$ (dotted curve).  In Eq.\,(\ref{pdf-L}) $120$ terms  were  used.
The solid curve shown in the last sub-figure is the stationary distribution, independent of $\alpha$. }
\label{Fig2}
\end{figure}


\onecolumngrid

\begin{figure}[ht] \centering
\includegraphics*[width=8cm,height=6cm]{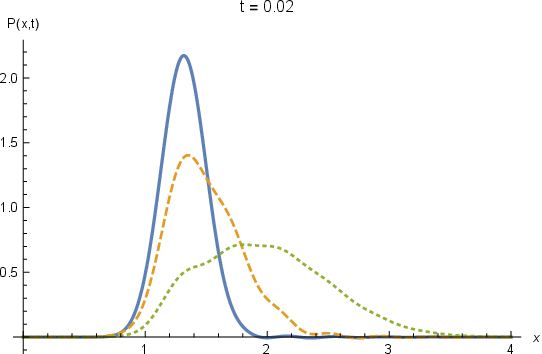}\hspace{1cm}
\includegraphics*[width=8cm,height=6cm]{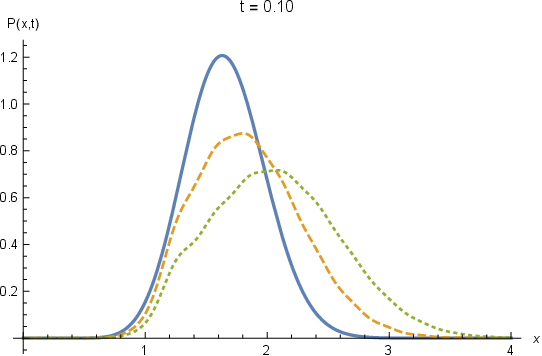}\\
~\\
~\\
~\\
\includegraphics*[width=8cm,height=6cm]{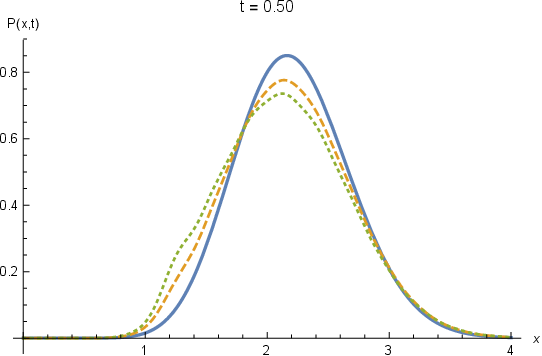}\hspace{1cm}
\includegraphics*[width=8cm,height=6cm]{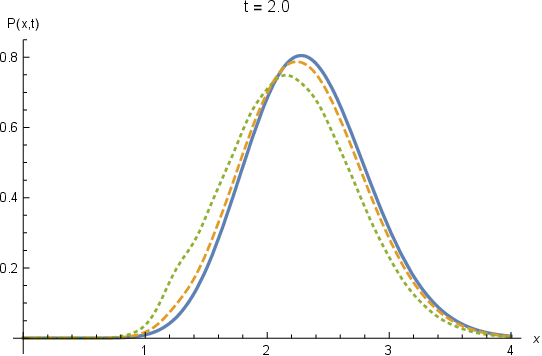}
~\\
~\\
~\\
\includegraphics*[width=8cm,height=6cm]{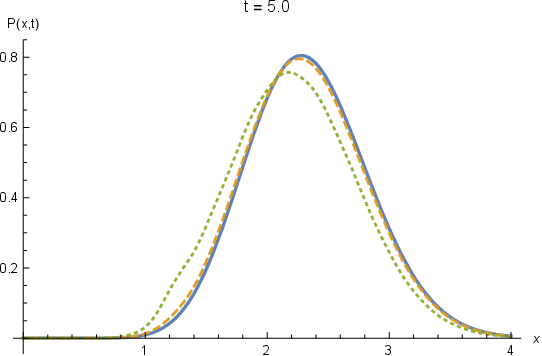}\hspace{1cm}
\includegraphics*[width=8cm,height=6cm]{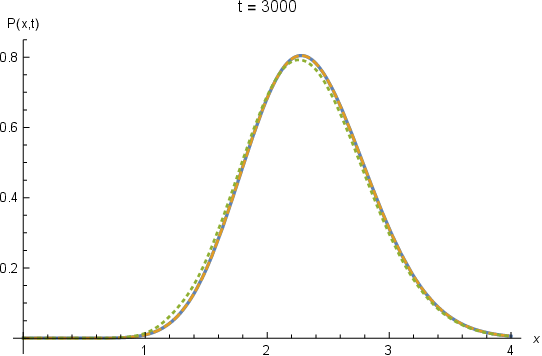}
\caption{Plots of $P(x,t)$ at different times for the $\ell$-deformed Rayleigh process  
with $g=0.5, \ell=5, x_0=1.2$, 
$\alpha=1$ (solid curve), $0.75$ (dashed curve) and $0.25$ (dotted curve).  In Eq.\,(\ref{pdf-L}) $120$ terms  were  used.
The solid curve shown in the last sub-figure is the stationary distribution, independent of $\alpha$.}

\label{Fig3}
\end{figure}

\end{document}